\documentclass[journal]{IEEEtran}
\usepackage{times,amsmath,amssymb,dsfont,graphicx,float,enumerate,cite,color}
\usepackage[linesnumbered,ruled,vlined]{algorithm2e}
\newtheorem{Theorem}{Theorem}

\newtheorem{Cor}{Corollary}
\newtheorem{Lemma}{Lemma}

\newcommand{\ket}[1]{\ensuremath{\left|#1\right\rangle}}
\DeclareMathOperator*{\argmin}{\arg\!\min}

\ifCLASSINFOpdf
\else
\fi
\IEEEoverridecommandlockouts

\begin{document}

\title{Subspace Selection for Projection Maximization with Matroid Constraints}
\author{Yuan~Wang, Zhenliang~Zhang,~\IEEEmembership{Member,~IEEE,}
        Edwin~K.~P.~Chong,~\IEEEmembership{Fellow,~IEEE,}\\
        Ali~Pezeshki,~\IEEEmembership{Member,~IEEE,}
        Louis~Scharf,~\IEEEmembership{Life Fellow,~IEEE}

\thanks{Y. Wang is with the Department of Biostatistics, University of Texas MD Anderson Cancer Center, Houston, TX 77030 USA
(e-mail: ywang46@mdanderson.edu). }

\thanks{Z.~Zhang is with Qualcomm Corporate R\&D, Bridgewater, NJ 08873, USA (e-mail: zhenlian@qti.qualcomm.com). He was with the Department of Electrical and Computer Engineering, Colorado State University, Fort Collins, CO 80523-1373, USA.}

\thanks{E.~K.~P.~Chong and A.~Pezeshki are with the Department of Electrical and Computer Engineering and Department of Mathematics, Colorado State University, Fort Collins, CO 80523-1373, USA (e-mail: edwin.chong@colostate.edu; ali.pezeshki@colostate.edu).}

\thanks{L.~L.~Scharf is with the Department of Mathematics and Department of Statistics, Colorado State University, Fort Collins, CO 80523, USA (e-mail: louis.scharf@colostate.edu).}
    }

\maketitle

\begin{abstract}

Suppose  that there is a ground set which consists of a large number of vectors in a Hilbert space. Consider the problem of selecting a subset of the ground set such that the projection of a vector of interest onto the subspace spanned by the vectors in the chosen subset reaches the maximum norm. This problem is generally NP-hard,  and alternative approximation algorithms such as \emph{forward regression} and \emph{orthogonal matching pursuit} have been proposed as heuristic approaches. In this paper, we investigate bounds on the performance of these algorithms by introducing the notions of \emph{elemental curvatures}. More specifically, we derive lower bounds, as functions of these elemental curvatures, for performance of the aforementioned algorithms with respect to that of the optimal solution under uniform and non-uniform \emph{matroid} constraints, respectively. We show that if the elements in the ground set are mutually orthogonal, then these algorithms are optimal when the matroid is uniform and they achieve at least $1/2$-approximations of the optimal solution when the matroid is non-uniform.
\end{abstract}

\section{Introduction}
Consider the Hilbert space $L^2(\mu)$ of square integrable random variables with $\mu$ the probability measure.  Let $X$ be a \emph{ground set} of vectors and $\eta$ be the vector of interest in $L^2({\mu})$.  Let $I$ be a non-empty collection of subsets of $X$, or equivalently, a subset of the power set $2^{X}$. For any set $E\in I$, we use $\text{span}(E)$ to denote the subspace spanned by the vectors in $E$. We use $\mathcal P_{\eta}(E)$ to denote the projection of $\eta$ onto $\text{span}(E)$. The goal is to choose an element $E$ in $I$ such that the square norm of $\mathcal P_{\eta}(E)$ is maximized, i.e.,
\begin{align}
\begin{array}{l}
\text{maximize  } \| \mathcal P_{\eta}(E)\|^2 \\
\text{subject to  } E\in I.
\end{array}
\label{obj}
\end{align}

\subsection{Motivating Examples}

The above formulation has vast applications in statistical signal processing \cite{scharf1991statistical}\cite{lehmann2003theory} such as
maximizing the quadratic covariance bound, sensor selection for minimizing the mean squared error, and sparse approximation for compressive sensing. Here we briefly introduce a few examples.
\begin{enumerate}
\item {Quadratic Covariance Bound.}

Let $\mu_\theta$ be the underlying probability measure associated with parameter $\theta$ lying on the parameter space $\Theta$. The problem of interest is to estimate $g(\theta)$, where $g: \Theta\to \mathbb R$ is a bounded known function. Let $\hat g \in L^2(\mu_{\theta})$ be an unbiased estimator of $g(\theta)$ and $\eta=\hat g-g(\theta) \in L^2(\mu_{\theta})$ be the estimation error, which is the vector of interest.
For any set $E$ of \emph{score functions}, the variance of any unbiased estimator is lower bounded by the square norm of the projection of estimation error $\eta$ onto $
\text{span}(E)$. This fact is also known as quadratic covariance bound\cite{mcwhorter1993properties},\cite{todros2010general}:
\begin{align}\label{Eqn:QCB}
\text{Variance}[\hat g]=\|\eta\|^2 \geq \| \mathcal P_{\eta}(E)\|^2,
\end{align}
where  $\|\eta\|^2={\mathbb E_{\mu_\theta}[\eta\eta^\ast]}$ and $\mathbb E_{\mu_\theta}$ denotes the expectation with respect to the measure $\mu_\theta$.
 The well-known Cramer-Rao bounds \cite{rao1945information}, Bhattacharyya bounds \cite{bh1947on}, and Barankin bounds \cite{barankin1946locally} are essentially special cases of the quadratic covariance bound by substituting $E$ with specific sets of score functions. For example, the score function for Cramer-Rao bounds is simply $\partial \ln d(x;\theta)/\partial \theta$, where $d(x;\theta)$ denotes the probability density function of measurement~$x$. While these established bounds provide insightful understandings for the performance of unbiased estimators, the corresponding score functions do not necessarily provide the tightest bounds for the estimator variance.
Moreover, derivation of these bounds such as Cramer-Rao bounds requires the inverse or pseudo-inverse \emph{Fisher information matrix}, which can be computationally impractical for large number/dimension of unknown parameters \cite{hero1997recursive}.  Last, a necessary condition to compute these bounds is that the probability density function and its partial derivatives are well-defined. For these reasons, other score functions might be more suitable for providing the lower bound. Suppose that there exists a large set $X$ of candidate score functions in $L^2(\mu_\theta)$. We aim to choose an optimal subset $E\subset X$ which maximizes $\| \mathcal P_{\eta}(E)\|^2$ and hence provides the tightest bound for variances of unbiased estimators.

\item{Linear Minimum Mean Squared Error Estimator.}

Suppose that there is a large set of sensors, each of which makes a zero-mean and square-integrable random sensor observation. These sensor observations are not necessarily independent. The goal is to select a subset of the sensors such that the mean squared error for estimating the parameter of interest $\eta$ is minimized. It is well-known that the orthogonality principle implies that the Linear Minimum Mean Squared Error (LMMSE) estimator, denoted by $\eta_{\text{LMMSE}}$, is the projection of $\eta$ onto the subspace spanned by a selected subset $E$ \cite{scharf1991statistical}. The problem of interest is how to choose $E$ from the set $X$ of all sensor observations such that the mean squared error $\mathbb E[(\eta_{\text{LMMSE}} -\eta)^2]$ is minimized, i.e., the projection of $\eta$ onto $\text{span}(E)$ is maximized. Another approach to this sensor selection problem is to maximize the information gain and apply submodularity to bound the performance of greedy algorithms \cite{williams2007information}-\nocite{krause2007near}\cite{shamaiah2010greedy}. When the criterion is mean squared error, the objective function is in general not submodular, resulting in difficulty to quantify the performance of the greedy algorithms.

\item{Sparse Approximation for Compressive Sensing.}

Compressive sensing is the problem of recovering a sparse signal using linear compressing measurements (see, e.g., \cite{candes2006compressive}--\nocite{donoho2006compressed,baraniuk2007compressive,eldar2012compressed,xu2007efficient}\cite{chi2011sensitivity}). Let $\eta\in \mathbb R^d$ be the measurement signal. We assume that $\eta=\mathbf H x$ where $\mathbf H \in \mathbb R^{d\times n}$ is the measurement matrix. The goal is to find $K$ non-zero components in the $n$-dimensional vector $x$ with $K\ll d<n$ such that  $\mathbf H x$ can exactly recover or well-approximate $\eta$, i.e.,
\begin{align*}
\begin{array}{l}
\text{minimize  }  \|\eta-\mathbf H x\|^2  \\
\text{subject to  } \|x\|_0\leq K,
\end{array}
\end{align*}
where $\|x\|_0$ denotes the $L_0$-norm of $x$.
The geometrical interpretation of the above problem is to select $K$ columns of matrix $\mathbf H$ such that the norm of the projection of $\eta$ onto the subspace spanned by the chosen columns is maximized. Adaptive algorithms such as those based on partially observable Markov decision processes have been proposed to find the optimal solution \cite{zahedi2013adaptive}. The computation complexity for adaptive algorithms is in general quite high despite the reduction brought by approximation methods such as rollout.
\end{enumerate}

All the above applications are special cases of the projection maximization problem defined in \eqref{obj}. In general, problem \eqref{obj} is a combinatorial optimization problem and it is NP-hard to obtain the optimal solution. Alternative algorithms such as forward regression~\cite{miller2002subset} and orthogonal matching pursuit~\cite{tropp2004greed}--\nocite{chi2012coherence,kunis2007random,cai2011orthognal}\cite{davenport2010analysis} have been studied intensively to approximate the optimal solution of \eqref{obj}. Each of these two algorithms starts with an empty set, and then incrementally adds one element to the current solution by optimizing a local criterion, while the updated solution still belongs to the set of feasible solutions $I$. They are known as greedy approaches due to the nature of local optimality, although the local criteria are different\footnote{Other variations of greedy approaches have also been proposed and investigated (see, e.g., \cite{braunnewinfo}\cite{liu2012the}).}. Details are given in Algorithms~\ref{alg1} and~\ref{alg2}, respectively. The definition of matroid will be given in Section~\ref{sec2}. Moreover, we use $\langle r | s \rangle$ to denote the inner product of $r$ and $s$ in the Hilbert space. Notice that neither algorithm achieves the maximum projection in general. The main purpose of this paper is to quantify their performance with respect to that of the optimal solution. We note that another frequently used approach is through convex relaxation schemes based on sparse-eigenvalue or restricted isometry property~\cite{candes2006stable}, although the objective there is usually to minimize the difference between the actual and estimated coefficients of sparse vectors (this corresponds to $L_0$-norm minimization while \eqref{obj} deals with $L_2$-norm).

\begin{algorithm}

 \SetAlgoLined
 \SetKwData{Left}{left}\SetKwData{This}{this}\SetKwData{Up}{up}
\SetKwFunction{Union}{Union}\SetKwFunction{FindCompress}{FindCompress}
\SetKwInOut{Input}{Input}\SetKwInOut{Output}{Output}

 \Input{Ground set $X$ and an associated matroid $(X, I)$; vector of interest $\eta$.}
 \Output{An element $E\in I$.}
\Begin{
 $E\leftarrow \emptyset$;

 \For{$\ell=1$ to $K$ }{

  $s^*=\underset {s\in X\setminus E,E\cup \{s\} \in I}{\arg\max} \|\mathcal P_{\eta}(E\cup\{s\})\|^2$;

  Update $E\leftarrow E\cup\{s^*\}$;
   }
 }
 \caption{Forward Regression}
\label{alg1}
\end{algorithm}

\begin{algorithm}
\label{alg2}
 \SetAlgoLined
 \SetKwData{Left}{left}\SetKwData{This}{this}\SetKwData{Up}{up}
\SetKwFunction{Union}{Union}\SetKwFunction{FindCompress}{FindCompress}
\SetKwInOut{Input}{Input}\SetKwInOut{Output}{Output}

 \Input{Ground set $X$ and  an associated matroid $(X, I)$; vector of interest $\eta$.}
 \Output{An element $E\in I$.}
\Begin{
 $E\leftarrow \emptyset$;

 Residue $r=\eta$;

 \For{$\ell=1$ to $K$ }{

  $s^*=\underset {s\in X\setminus E , E\cup \{s\} \in I}{\arg\max} | \langle r | s \rangle|$;

  Update $E\leftarrow E\cup\{s^*\}$;

  Update $r \leftarrow  r - \mathcal P_{ \eta}(E)$;
   }
 }
 \caption{Orthogonal Matching Pursuit}
\end{algorithm}

\subsection{Main Contributions}
The main purpose of this paper is to provide performance bounds for forward regression and orthogonal matching pursuit with respect to the optimal solution.
To derive the bounds, we will define several notions of elemental curvatures, which are inspired by the elemental curvature introduced in \cite{wang2012}. We also illustrate from a geometric perspective how these elemental curvatures are related with principal angles, which are in turn related with the {restricted isometry property} and {mutual incoherence} \cite{donoho2006stable}. It turns out that the (near-)optimality of the two aforementioned algorithms is closely related with the mutual (near-)orthogonality of the vectors in the ground set and the structure of the matroid. Our approach allows the derivation of sharp approximation bounds for these two algorithms, in general situations (where the matroid might be uniform or non-uniform). To the best of our knowledge, the non-uniform matroid situation has never been investigated in any previous papers. More specifically, in the special case where the vectors in the ground set are mutually orthogonal, these two algorithms are optimal when the matroid is uniform and they achieve at least $1/2$-approximations of the optimal solution when the matroid is non-uniform.

\section{Curvatures, Matroid, and Related Work}\label{sec2}
In this section, we first introduce two new notions of curvature and review the definition the matroid. Then we review the related literature to our study. Last, we investigate the notions of curvature from a geometric perspective. 

As we shall see later from this geometric perspective, curvatures are essentially metrics to capture the mutual near-orthogonality of the vectors in the ground set. Without loss of generality, throughout the paper we assume that all elements in $X$ are normalized, i.e., $\|t\|^2 =1$ for any $t\in X$. Let $t^\perp(E)$ and $\bar t(E)$ be the normalized orthogonal and parallel components of $t$ with respect to $\text{span}(E)$ (simplified as $t^{\perp}$ and $\bar t$ unless otherwise specified):
\[
t=t^{\perp} \sin\varphi + \bar t\cos\varphi,
\]
where $\varphi$ denotes the angle between $t$ and $\text{span}(E)$.

We define the \textbf{\emph{forward elemental curvature}}, denoted by $\hat\kappa$, as follows:
\begin{align*}
\hat  \kappa &=\max_{E, s, t}\frac{\|\mathcal P_\eta(E\cup\{s,t\})\|^2-\|\mathcal P_\eta(E\cup\{s\})\|^2}{\|\mathcal P_\eta(E\cup\{t\})\|^2-\|\mathcal P_\eta(E)\|^2}\\
&\text{subject to } E\subset X,  s,t\in X\setminus E, \text{card}(E)\leq 2K-2, \\
&\text{and }\mathcal \|P_{\eta}(\{s^{\perp}(E)\})\| \leq \|\mathcal P_{\eta}(\{t^{\perp}(E)\})\| .
\label{curvature}
\end{align*}
Similarly, we define the \textbf{\emph{backward elemental curvature}}, denoted by $\bar \kappa$ as follows:
\begin{align*}
 \bar \kappa=&\max_{E, s, t} \frac{\|\mathcal P_\eta(E\cup\{s,t\})\|^2-\|\mathcal P_\eta(E\cup\{s\})\|^2}{\|\mathcal P_\eta(E\cup\{s\})\|^2-\|\mathcal P_\eta(E)\|^2}\\
& \text{subject to } E\subset X, \text{card}(E)\leq 2K-2, s,t\in X\setminus E, \\
 & \text{and }\mathcal \|P_{\eta}(\{s^{\perp}(E)\})\| \geq \|\mathcal P_{\eta}(\{t^{\perp}(E)\})\|.
\end{align*}
Notice that both curvatures are ratios of \emph{differences} of the discrete function, analogous to second-order derivative of a continuous function. In particular, if all the elements in $X$ are mutually orthogonal, then $\hat\kappa = \bar\kappa = 1$.
Moreover, it is easy to show that the objective function in \eqref{obj} is always monotone:
Suppose that $S\subset T\subset X$. Then, by definition, $\text{span}(S)$ is a subspace of $\text{span}(T)$. Thus we have
\[\|\mathcal P_{\eta}(S)\|^2 \leq \|\mathcal P_{\eta}(T)\|^2,
\]
which indicates that $\hat \kappa$ and $\bar \kappa$ are always non-negative.

Next we state the definition of  \emph{\textbf{matroid}}. Let $I$ be a collection of subsets of $X$.  We call $(X,I)$ a matroid~\cite{tutte1965lectures} if it has the \emph{hereditary}  property:  For any $S\subset T\subset X$, $T\in  I$ implies that $S\in  I$; and the \emph{augmentation} property: For any $S, T \in  I$, if $T$ has a larger cardinality than $S$, then there exists $j\in T\setminus S$ such that $S\cup \{j\}\in  I$. Furthermore, we call $(X, I )$ a \emph{uniform} matroid if  $ I=\{S\subset X: \text{card}(S)\leq K\}$ for a given $K$, where $\text{card}(S)$ denotes the cardinality of $S$. Otherwise, $(X, I)$ is a \emph{non-uniform} matroid. The structure of a matroid captures the feasible combinatorial solutions within the power set of the ground set. Take the sensor selection problem as an example, a uniform matroid constraint means that we can choose any combination of $K$ sensors from all the sensors for the solution; a non-uniform matroid constraint means that only certain combinations of $K$ sensors are feasible solutions. Similarly, in many compressed sensing applications such as \cite{zhang2013neighbor}, we might have some prior knowledge that not all combinations of sparsity locations are feasible solutions.

\subsection{Related Work}\label{sec12}
We first review the notion of submodular set function. Let $X$ be a ground set and $f:2^X\to \mathbb R$ be a function defined on the power set $2^X$. We call that $f$ is submodular if
\begin{itemize}
\item[1)] $f$ is non-decreasing: $f(A)\leq f(B)$ for all $A\subset B$;
\item [2)] $f(\emptyset)=0$ where $\emptyset$ denotes the empty set (note that we can always substitute $f$ by $f-f(\emptyset)$ if $f(\emptyset)\neq 0$);
     \item [3)] $f$ has the diminishing-return property: For all $A\subset B\subset X$ and $j\in X\setminus B$, we have $f(A\cup \{j\})-f(A)\geq f(B\cup \{j\})-f(B)$.
\end{itemize}
The optimization problem that aims to find a set in the matroid to maximize a submodular function is in general not tractable. Many papers have studied the greedy algorithm as an alternative: starting with an empty set, incrementally add one more element that maximizes the local gain of the objective function to the current solution, while the updated solution still lies in the matroid. Existing studies have shown that the greedy algorithm approximates the optimal solution well. More specifically,
Nemhauser \emph{et al.}~\cite{nemhauser1978analysis} showed that the greedy algorithm achieves at least a $(1-e^{-1})$-approximation for a uniform matroid. Fisher \emph{et al.} \cite{fisher1978analysis} proved that the greedy algorithm provides at least a $1/2$-approximation of the optimal solution for a non-uniform matroid. Moreover, let $\kappa_t$ be the total curvature of function $f$,  which is defined as   \[
\kappa_t=\max_{j\in X}\left\{1-\frac{f(X)-f(X\setminus \{j\})}{f(\{j\})-f(\emptyset)}\right\}.
\] Conforti and Cornu\'{e}jols \cite{conforti1984submodular} showed that the greedy algorithm achieves at least $\frac{1}{\kappa_t}(1-e^{-\kappa_t})$ and $\frac{1}{1+\kappa_t}$-approximations of the optimal solution for uniform and non-uniform matroids, respectively. Note that $\kappa_t\in [0,1]$ for any submodular function, and the greedy algorithm is optimal when $\kappa_t=0$.
Vondr\'{a}k \cite{vondrak2010submodularity} showed that the \emph{continuous greedy algorithm} achieves at least a $\frac{1}{\kappa_t}(1-e^{-\kappa_t})$-approximation for any matroid. On the other hand, Wang~\emph{et~al.} \cite{wang2012} provided approximation bounds for the greedy algorithm as a function of elemental curvature, which generalizes the notion of diminishing return and is defined as
\[
\kappa_e=\max_{E\subset X, i, j \in X\setminus E, i\neq j} \frac{f(E\cup \{i,j\}) -f(E\cup \{j\})}{f(E\cup \{i\}) -f(E)}.
\]
Note that the objective function is submodular if and only if $\kappa_e\leq 1$. When $\kappa_e<1$, the lower bound for greedy approximation is greater than $(1-e^{-1})$. If $\kappa_e>1$, then the objective function is not submodular. In this case, lower bound for the greedy algorithm is derived as a function of the elemental curvature. In \cite{zhang2013cdc} and \cite{zhang2013near}, Zhang~\emph{et~al.} generalized the notions of total curvature and elemental curvature to \emph{string submodular functions} where the objective function value depends on the order of the elements in the set. This framework is further extended to approximate dynamic programming problems by Liu~\emph{et al.} in \cite{liu2014cdc}.

We use $\ket i$ to denote the orthonormal bases of the Hilbert space, $i=0,1,\ldots\,$. The objective function in \eqref{obj} is not submodular in general.  For example, let $\eta=\ket{0}$, $s=\ket 1$, and $t=\frac{1}{2}\ket 0+\frac{\sqrt 3}{2}\ket 1$. Then we have
\begin{align*}
&\|\mathcal P_\eta(\{t\})\|^2-\|\mathcal P_\eta(\emptyset)\|^2 =\frac{1}{4};\\
&\|\mathcal P_\eta(\{s,t\})\|^2-\|\mathcal P_\eta(\{s\})\|^2 =1>\frac{1}{4}.
\end{align*}
Evidently the diminishing return property does not hold in this case.
In fact, the diminishing return property does not always hold even if all the elements in the ground set are mutually orthogonal. Therefore, the results from classical submodularity theory~(e.g.,\cite{nemhauser1978analysis}\cite{fisher1978analysis}) are not directly applicable to our problem. To address this issue, several notions of \emph{approximation submodularity} are introduced to bound the greedy algorithm performance. Cevher and Krause \cite{cevher2011greedy} showed that the greedy algorithm achieves a good approximation for sparse approximation problems using the approach of {approximation submodularity}. Das and Kempe \cite{das2011} improved the approximation bound by introducing the notion of \emph{submodularity ratio}. These are powerful results, but with limited extension to non-uniform matroid structures. In this paper,  we will use the aforementioned notions of curvature to bound the performance of forward regression and orthogonal matching pursuit with respect to the optimal solution even if the matroid is non-uniform.

\subsection{Geometric Interpretation of Curvatures}

To understand the curvatures from a geometric perspective,
we define the \emph{principal angle} as follows:
\[
\phi =\min_{E\subset X, |E|\leq 2K-2, s\in X\setminus E}\arccos \|\mathcal P_{s}(E)\|,
\]
where $\phi\in[0,\pi/2]$.
Geometrically speaking, this is saying that the angle between the subspace spanned by any subset $E$ (with cardinality less than or equal to $2K-2$) and any element in the set $X\setminus E$  is not smaller than $\phi.$ Note that if all the elements in $X$ are mutually orthogonal, then $\phi=\pi/2$.

We now investigate the relationship between the principal angle and two widely used conditions in compressed sensing to quantify the performance of recovery algorithms, namely {restricted isometry} and {mutual incoherence}. Let $\mathbf H=[h_1,h_2,\ldots,h_m]$ be the matrix associated with $E=\{h_1,h_2,\ldots,h_m\}$. It is easy to see that
\begin{align*}
\cos\phi&=\max_{E\subset X, |E|\leq 2K-2,s\in X\setminus E}\| \mathcal P_{s}(E)\|\\
&=\max_{E\subset X, |E|\leq 2K-2, s\in X\setminus E}\|\mathbf H(\mathbf H^T\mathbf H)^{-1}\mathbf H^T s\| \\
&\leq \max_{E\subset X, |E|\leq 2K-2, s\in X\setminus E} \|\mathbf H(\mathbf H^T\mathbf H)^{-1} \| \|\mathbf H^T s\|.
\end{align*}
The last inequality is by the Cauchy-Schwarz inequality. Moreover, we have
\begin{align*}
\|\mathbf H(\mathbf H^T\mathbf H)^{-1} \| &=\sup_{\|x\|=1} \|\mathbf H(\mathbf H^T\mathbf H)^{-1} x\| \\
&=\sqrt{\lambda_{\max}((\mathbf H(\mathbf H^T\mathbf H)^{-1})^T \mathbf H(\mathbf H^T\mathbf H)^{-1}}\\
&=\sqrt{\lambda_{\max}(\mathbf H^T\mathbf H)^{-1} } \\
&=\left(\sqrt{\lambda_{\min} (\mathbf H^T \mathbf H)}\right)^{-1}.
\end{align*}
and
\begin{align*}
\|\mathbf H^T s\|=\left( \sum_{i=1}^m \langle h_i| s \rangle^2 \right)^{\frac{1}{2}}.
\end{align*}
Thus, we have
\begin{align}\label{angleRIP}
&\cos \phi\leq\\ \nonumber
&\max_{E\subset X, |E|\leq 2K-2, s\in X\setminus E} \left(\sqrt{\lambda_{\min} (\mathbf H^T \mathbf H)}\right)^{-1} \left( \sum_{i=1}^m \langle h_i| s \rangle^2 \right)^{\frac{1}{2}}.
\end{align}
Here $\lambda_{\min}(\mathbf H^T\mathbf H)$ denotes the minimum eigenvalue of the correlation matrix $\mathbf H^T\mathbf H$, which is closely related with the restricted isometry property. The summation term for the inner products is upper bounded by $m$ times the squared mutual incoherence.

Next we present a result that bridges curvatures and principal angle.
\begin{Theorem} \label{thm2}
Forward and backward elemental curvatures are both upper bounded as:
\[
\max(\hat\kappa, \bar\kappa) \leq \frac{1}{1-2\cos\phi}.
\]

\end{Theorem}

The proof is given in Appendix~\ref{app1}. This result is important in the cases where the curvatures are difficult to calculate. We can use the principal angle, or an upper bound for the principal angle such as \eqref{angleRIP} to bound the curvature, which in turn provides performance bounds for forward regression and orthogonal matching pursuit. 

Next we study the performance of forward regression and orthogonal matching pursuit with uniform and non-uniform matroid constraints.
We will use $f(E)$ to represent $\|\mathcal P_{\eta}(E)\|^2$ occasionally in the following sections to simplify notation.

\section{Results for uniform Matroid}
In this section, we will focus on the case where the matroid is uniform, i.e.,  $I=\{S\subset X: \text{card}(S)\leq K\}$ for a given $K$. We consider two scenarios depending on the mutual orthogonality of elements in $X$.

\subsection{Orthogonal Scenario}
 We call the set $X$ mutually orthogonal if any two non-identical elements in $X$ are orthogonal: $\langle s| t \rangle =0$ for any $s\neq t \in X$. It is easy to show that forward regression and orthogonal matching pursuit are equivalent given that $X$ is mutually orthogonal.
It turns out that the optimality of these two algorithms is closely related with the mutual orthogonality of $X$.
\begin{Theorem} Suppose that $X$ is mutually orthogonal. If $(X,I)$ is a uniform matroid, then forward regression and orthogonal matching pursuit are optimal.
\label{thm1}
\end{Theorem}
\begin{IEEEproof}
Let $E=\{e_1,\ldots,e_K\}$ be a subset and $\eta$ be the vector of interest. By the Hilbert projection theorem and Pythagoras' theorem, we have
\[
\|\mathcal P_{\eta}(E)\|^2 ={\sum_{i=1}^K \langle \eta| e_i \rangle ^2}.
\]
It is easy to see that the optimal solution is to choose $K$ largest projections among all vectors in $X$, which is the same as what the forward regression does. The insight of this result is closely related with \emph{principle component analysis}.
\end{IEEEproof}

Theorem~\ref{thm1} implies that to guarantee the optimality of forward regression and orthogonal matching pursuit, we should find an orthonormal basis for $X$. The Gram--Schmidt process can be used to generate an orthonormal basis using the elements in $X$. However, this is, in general, intractable especially when $\text{card}(X)$ is large. Moreover, the problem of optimally selecting $K$ elements in $X$ is different from the problem of optimally selecting $K$ orthogonalized elements after applying the Gram--Schmidt process.

Mutual orthogonality depends on the definition of inner product in the Hilbert space. For example, the Hilbert space defined on Gaussian measures has an orthonormal basis: Hermite polynomials. Some other well-known examples include Charlier polynomials for Poisson measures, Laguerre polynomials for Gamma measures, Legendre and Fourier polynomials for uniform measures.

The physical meaning of mutual orthogonality differs depending on the context of the problem. Take the quadratic covariance bound problem for example and consider the uniform distribution parameterized by its mean $\theta$: Uniform$[-\pi+\theta,\pi+\theta]$. The Cramer-Rao Bound is not applicable here because the derivative of the probability density function is not well-defined. On the other hand, the Fourier basis $\{\cos(m(x-\theta))\}_{m=1}^{\infty}$ is a well-defined orthonormal basis. These basis functions can be considered as energy eigenstates for a quantum particle in an infinite potential well.
Another example is the Bhattacharya bound with the following Bhattacharya score functions:
\[
\left\{\frac{\partial \ln d(x,\theta)}{\partial \phi}, \frac{\partial^2 \ln d(x,\theta)}{\partial \phi^2},\ldots,  \frac{\ln \partial^k d(x,\theta)}{\partial \phi^k} \ldots \right\},
\]
where $d(x,\theta)$ denotes the probability density function for the measurement $x$. In general, these score functions are not orthonormal. Moreover, the projection of the estimator error onto the first order partial derivative is not necessarily the largest, meaning that the Fisher score is not necessarily the optimal. However, in the Gaussian measure case, the Bhattacharya score functions turn out to be the Hermit polynomials and therefore are mutually orthogonal.
 For the LMMSE problem, mutually orthogonality means that all the sensor measurements are mutually \emph{uncorrelated}. Therefore, if all the sensors generate independent measurement signals, then forward regression and orthogonal matching pursuit are optimal in the uniform matroid case. For the sparse approximation problem, mutual orthogonality says that all the columns in the measurement matrix are mutually orthogonal, which cannot be true in the case of the under-determined system.

\subsection{Non-orthogonal Scenario}
When $X$ is not mutually orthogonal, forward regression and orthogonal matching pursuit are  in general  not optimal. We give a counter example for forward regression; a similar counter example can be given for orthogonal matching pursuit.  Let $X=\{s_1,s_2,s_3\}$ where $s_1=\frac{\sqrt 2}{2}(\ket{0}+\ket{1})$, $s_2=\frac{\sqrt 2}{2}(\ket{1}+\ket{2})$, and $s_3=\frac{\sqrt 2}{2}(\ket{2}+\ket{3})$. Suppose that ${\eta}=\ket{0}+2\ket{1}+2\ket{2}+\ket{3}$, and the objective is to choose a subset $E$ of $X$ with $\text{card}(E)\leq 2$ such that the projection of $\eta$ onto $\text{span}(E)$ is maximized. Obviously, the optimal solution is to choose $s_1$ and $s_3$ and the maximum projection is
\[
\|\mathcal P_{\eta}(\{s_1,s_3\})\|^2={ \langle \eta|s_1 \rangle ^2 +\langle \eta| s_3 \rangle ^2 }= 9.
\]
Forward regression, however, is fooled into picking $s_2$ first because along $s_2$ it has the largest projection. After that, it chooses either $s_1$ or $s_3$.  By the Gram--Schmidt process, the normalized orthogonal component of $s_1$ with respect to $s_2$ is given by
\[
{s}_1^\perp = \frac{ s- \langle s_1| s_2 \rangle s_2} {\|s_1- \langle s_1| s_2 \rangle s_2\|} =\sqrt \frac{4}{3} (\frac{\sqrt 2}{2} \ket 0 +\frac{\sqrt 2}{4} \ket 1-\frac{\sqrt 2}{4} \ket 2).
\]
Therefore,
\begin{align*}
&\|\mathcal P_{\eta}(\{s_1,s_2\})\|^2=\|\mathcal P_{\eta}(\{ s_1^\perp,s_2\})\|^2 \\
&={ \langle \eta| s_2 \rangle ^2 +\langle \eta|  s_1^\perp \rangle ^2} = {8+\frac{2}{3}}.
\end{align*}
Apparently, forward regression is not optimal. Moreover, if $X$ is not mutually orthogonal, then the two algorithms yield different results, which we discuss in separate subsections.

\subsubsection{Forward Regression}
We first study forward regression when the matroid is uniform with the maximal cardinality of the sets in $I$ equal to $K$. We use $G_K$ to denote the solution using forward regression and $\text{OPT}$ to denote the optimal solution.
\begin{Theorem}[Uniform matroid] The forward regression algorithm achieves at least a $(1-(1-\frac{1}{\hat K})^K)$-approximation of the optimal solution:
\begin{align}
\label{eqthm3}
f(G_K) \geq \left(1-\left(1-\frac{1}{\hat K}\right)^K\right)f(\text{OPT}),
\end{align} where $\hat K=\sum_{i=1}^{K} \min(\bar \kappa,\hat \kappa)^{i-1}$.
\label{thm3}
\end{Theorem}

The proof is given in Appendix~\ref{app2}. When $\min(\bar \kappa, \hat\kappa) \leq 1$, the forward regression algorithm achieves at least a $(1-1/e)$-approximation of the optimal solution.

\subsubsection{Orthogonal Matching Pursuit}
We first compare the step-wise gains in the objective function between orthogonal matching pursuit and forward regression. Recall that $\eta^\perp$ and $\bar \eta$ represent the normalized orthogonal and parallel components of $\eta$ with respect to $\text{span}(E)$:
\[
\eta=\eta^{\perp} \sin\varphi + \bar\eta\cos\varphi,
\]
where $\varphi$ denotes the angle between $\eta$ and $\text{span}(E)$.   The orthogonal matching pursuit algorithm aims to find an element $t$ to maximize $|\langle \eta^\perp| t\rangle|$. The forward regression algorithm aims to find an element $s$ to maximize $|\langle \eta| s^{\perp}\rangle|$, where $s^\perp$ denotes the normalized orthogonal component of $s$ with respect to $\text{span}(E)$. Suppose that the angle between  $s$ and $\text{span}(E)$ is $\delta(s)$. Note that $\delta(s)$ is lower bounded by the principal angle $\phi$ by definition. By the fact that
\begin{align}
\label{eqcomp}
\nonumber
&\max_{s\in X\setminus E} \langle \eta^{\perp}| s \rangle^2\\
&=\max_{s\in X\setminus E} \langle \eta^{\perp}| s^{\perp} \sin \delta(s) \rangle^2\\
\nonumber
&\geq \sin^2 \phi \max_{s\in X\setminus E} \langle \eta^{\perp}| s^{\perp} \rangle^2\\
\nonumber
&\geq \sin^2 \phi \max_{s\in X\setminus E} \langle \eta| s^{\perp} \rangle^2,
\end{align}
even though orthogonal matching pursuit is not the ``greediest'' algorithm, its step-wise gain is still within a certain range of that of forward regression, captured by the principal angle. With this observation, we can derive a performance bound for orthogonal matching pursuit. Again, we assume that the matroid is uniform with the maximal cardinality of the sets in $I$ equal to $K$. We use $T_K$ to denote the solution using orthogonal matching pursuit.

\begin{Theorem}[Uniform matroid] \label{thm5}
The orthogonal matching pursuit algorithm achieves at least a $(1-(1-\frac{\sin^2 \phi}{\hat K})^K)$-approximation of the optimal solution:
\begin{align}
\label{eqthm4}
f(T_K) \geq \left(1-\left(1-\frac{\sin^2 \phi}{\hat K}\right)^K\right)f(\text{OPT}),
\end{align}
where $\hat K=\sum_{i=1}^{K} \min(\bar \kappa,\hat \kappa)^{i-1}$.
\end{Theorem}

The proof is given in Appendix~\ref{app4}. Notice that the difference between Theorem~\ref{thm3} and Theorem~\ref{thm5} is only the principal angle term $\sin^2\phi$. It is easy to see that the lower bound in \eqref{eqthm4} is always lower than that in \eqref{eqthm3}, but this does not necessarily mean that $f(T_k) \leq f(G_k)$.

\section{Results for non-uniform matroid} \label{sec:exp}
For non-uniform matroids, the two algorithms are not necessarily optimal even when $X$ is mutually orthogonal. As a counter example, suppose that $X=\{\ket 0, \ket 1, \ket 2, \ket 3\}$ and $I= \{\{\ket 0\}, \{\ket 1\},\{\ket 2\}, $ $ \{\ket 3\}, \{\ket 0, \ket 1\}, \{\ket 2, \ket 3\}\}$. It is easy to verify that $(X, I)$ is a non-uniform matroid. Let $\eta = \sqrt{1+\epsilon} \ket 0 + \ket 2 +\ket 3$ be the vector of interest, where $\epsilon>0$. Forward regression ends up with $\{\ket 0, \ket 1\}$ while the optimal solution is $\{\ket 2, \ket 3\}$. However, notice that \[\frac{\|\mathcal P_\eta(\{\ket 0, \ket 1\})\|^2}{\| \mathcal P_\eta(\{\ket 2, \ket 3\})\|^2} = (1+\epsilon) /2 > 1/2.\] In this section, we will show that $1/2$ is a general lower bound of these two algorithms for the non-uniform matroid case when the ground set is mutually orthogonal. This bound surprisingly matches the bound in \cite{fisher1978analysis}. However, a significant distinction is that  in our paper the submodularity of the objective function is no longer necessary (which is required by \cite{fisher1978analysis}).

Next we derive performance bounds for forward regression and orthogonal matching pursuit in the situation where $(X, I)$ is a non-uniform matroid.
Before proceeding, we state a lemma that assists in handling the non-uniform matroid constraint. Let $G_i$ and $T_i$ be the forward regression and orthogonal matching pursuit solutions up to step $i$, respectively. Note that the cardinalities of $G_i$ and $T_i$ are $i$.

\begin{Lemma} \label{lem1}
 Any $E\subset X$ with cardinality $K$ can be ordered into $\{e_1,\ldots, e_K\}$ such that for $i=1,\ldots, K$, we have
\[
f(G_{i-1}\cup \{e_i\})-f(G_{i-1}) \leq f(G_{i})-f(G_{i-1})
\]
and
\[
f(T_{i-1}\cup \{e_i\})-f(T_{i-1}) \leq f(T_{i-1}\cup \{g^*\})-f(T_{i-1}),
\]
where $g^*$ denotes the element added to $T_{i-1}$ using the forward regression algorithm.
\label{lemma}
\end{Lemma}
\begin{IEEEproof}
We prove this lemma using induction in descending order on the index $i$. First consider the sets $E$ and $G_{K-1}$, and notice that $|E|=K> |G_{K-1}|$. By the augmentation property of matroids, there exists an element in $E$, denoted by $e_K$, such that $G_{K-1}\cup\{e_K\} \in I$. It is easy to see that $f(G_K)-f(G_{K-1}) \geq f(G_{K-1}\cup\{e_K\}) -f(G_{K-1})$. Suppose that $f(G_k)-f(G_{k-1}) \geq f(G_{k-1}\cup\{e_k\}) -f(G_{k-1})$ for all $k \geq i$; we want to show that the inequality holds for the index $i-1$. Consider $G_{i-2}$ and $E\setminus \{e_i,\ldots, e_K\}$, where $e_k$ denotes the element in $E$ such that the claim holds for $k=i,\ldots, K$. Again by the augmentation property of matroids, there exists an element in $E\setminus \{e_i,\ldots, e_K\}$, denoted by $e_{i-1}$, such that $G_{i-2}\cup \{e_{i-1}\}\in I$. By the property of the forward regression algorithm, we know that $f(G_{i-1})-f(G_{i-2}) \geq f(G_{i-2}\cup\{e_{i-1}\}) -f(G_{i-2})$. This concludes the induction proof.

The proof for the orthogonal matching pursuit follows a similar argument and it is omitted for the sake of brevity.
\end{IEEEproof}

\subsubsection{Forward Regression}
In this section, we state the result for forward regression with the non-uniform matroid constraint. We first state a lemma.
\begin{Lemma}\label{lem2}
 For $i=1,2,\ldots, K$, we have
\[f(G_i) -f(G_{i-1}) \leq \bar \kappa (f(G_{i-1}) -f(G_{i-2})).
\]
\end{Lemma}
\begin{IEEEproof}
Let $G_i=\{g_1,\ldots, g_i\}$ where $g_j$ denotes the element added in the forward regression algorithm at step $j$.
We know that $G_{i-2}\cup \{g_{i}\} \in I$ because of the hereditary property of the matroid. Moreover, the projection of $\eta$ gains more by adding $g_{i-1}$ than $g_i$ at stage $i-1$ by the property of the forward regression algorithm. Then, by the definition of the backward elemental curvature, we obtain the desired result.
\end{IEEEproof}

Next we present the performance bound for forward regression in the non-uniform matroid scenario.

\begin{Theorem}[Non-uniform matroid] \label{thm4}
The forward regression algorithm achieves at least a $\frac{1}{1+a(\hat \kappa,\bar \kappa)b(\hat \kappa)}$-approximation of the optimal solution:
\[
f(G_K)\geq \frac{1}{1+a(\hat \kappa,\bar \kappa)b(\hat \kappa)} f(\text{OPT}),
\]
 where $a(\hat \kappa, \bar \kappa)=\max(\hat \kappa, \bar \kappa)$ if $\max(\hat \kappa, \bar \kappa) \leq 1$ and $a(\hat \kappa, \bar \kappa)=\max(\hat \kappa, \bar \kappa)^{K}$ if $\max(\hat \kappa, \bar \kappa) >1$; $b(\hat \kappa)=\hat \kappa^{K-1}$ if $\hat \kappa >1$ and $b(\hat \kappa)=1$ if  $\hat \kappa \leq 1$.
\end{Theorem}

The proof is given in Appendix~\ref{app3}.

\subsubsection{Orthogonal Matching Pursuit}
Next we derive the bound for orthogonal matching pursuit for the case where $(X,I)$ is a non-uniform matroid. To do so, we first define the OMP elemental curvature as follows:
\begin{align*}
\tilde{\kappa}=&\max_{E, s,t}  \frac{\|\mathcal P_\eta(E\cup\{s,t\})\|^2-\|\mathcal P_\eta(E\cup\{s\})\|^2}{\|\mathcal P_\eta(E\cup\{s\})\|^2-\|\mathcal P_\eta(E)\|^2}. \\
& \text{subject to } E\subset X, \text{card}(E)\leq 2K-2, s,t\in X\setminus E, \\
 & \text{and }\mathcal |\langle \eta^\perp(E)| s\rangle |\geq |\langle \eta^\perp(E)| t\rangle |.
\end{align*}
Again, we can provide an upper bound for OMP elemental curvature using principal angles. Note that $ |\langle \eta^\perp| s\rangle |\geq |\langle \eta^\perp| t\rangle |$ implies that
\[
\frac{|\langle \eta| t^\perp \rangle |}{|\langle \eta| s^\perp \rangle |} = \frac{|\langle \eta^\perp| t^\perp \rangle |}{|\langle \eta^\perp| s^\perp \rangle |} \leq \frac{1}{\sin \phi}.
\]
We can show that $\tilde\kappa$ is upper bounded as
\[
\tilde\kappa \leq \frac{(\sin^{-2} \phi + \langle t^{\perp}| s^\perp \rangle)^2 }{1-\langle t^{\perp}| s^\perp \rangle^2}.
\]
Similar to the technique in Theorem~\ref{thm2}, we can further bound the curvature using \eqref{ine}.
Next we state our result in the non-uniform matroid case.
\begin{Theorem}[Non-uniform matroid] \label{thm6}
  The orthogonal matching pursuit achieves at least a $\frac{1}{1+a(\hat \kappa,\bar \kappa,\tilde\eta)b(\hat \kappa){\sin^{-2}\phi}}$-approximation of the optimal solution:
  \[
f(T_K)\geq \frac{1}{1+a(\hat \kappa,\bar \kappa,\tilde\eta)b(\hat \kappa){\sin^{-2}\phi}}f(\text{OPT}),
\]
 where $a(\hat \kappa, \bar \kappa,\tilde \kappa)=\max(\hat \kappa, \bar \kappa ,\tilde \kappa)$ if $\max(\hat \kappa, \bar \kappa, \tilde \kappa) \leq 1$ and $a(\hat \kappa, \bar \kappa,\tilde \kappa)=\max(\hat \kappa, \bar \kappa,\tilde \kappa)^{K}$ if $\max(\hat \kappa, \bar\kappa,\tilde \kappa) >1$; $b(\hat \kappa)=\hat \kappa^{K-1}$ if $\hat \kappa >1$ and $b(\hat \kappa)=1$ otherwise.
\end{Theorem}

The proof is given in Appendix \ref{app5}. Note that when $X$ is mutually orthogonal, $\sin \phi=\max(\hat \kappa, \bar \kappa, \tilde\kappa) =1$. An immediate result follows.
\begin{Cor} Suppose that $X$ is mutually orthogonal. Then,
\begin{itemize}
\item[1)] Forward regression is equivalent to orthogonal matching pursuit;
\item[2)] If $I$ is a non-uniform matroid, then forward regression achieves at least a $1/2$-approximation of the optimal solution.
\end{itemize}
\end{Cor}

Recall that when $X$ is mutually orthogonal, we have shown in Section II-A that these two algorithms are optimal when $(X, I)$ is a uniform matroid. For a non-uniform matroid, they are not necessarily optimal. However, these two algorithms achieve at least $1/2$-approximations of the optimal solution. Our results extend those in \cite{fisher1978analysis}  from a submodular function to a more general class of objective functions.

Suppose that $X$ is not mutually orthogonal but close in the sense that the principal angle $\phi$ almost equal to $\pi/2$.
We use $\delta=\pi/2-\phi$ to denote the gap between $\phi$ and $\pi/2$. Moreover, we assume that $\delta$ is sufficiently small such that we only have to keep first order terms for Taylor expansions:
\[
\frac{1}{1-2\cos(\pi/2-\delta)} \approx 1+2\delta,
\]
 and
 \[
 \hat\kappa^{K-1} \approx 1+(K-1)|\hat\kappa-1|.
 \]Then, in the case of non-uniform matroid constraints, the lower bounds in Theorems~\ref{thm4} and~\ref{thm6} for the aforementioned algorithms scale as
\[
\frac{1}{2+2(2K-1)\delta},
\]
which indicates that the lower bound scales inverse linearly with cardinality constraints $K$ and the principal angle gap $\delta$ with $\pi/2$. Fortunately, $K$ is mostly a small number (for example, the number of sparsity locations in compressive sensing problem).

\section{Conclusions}
In this paper, we have studied the subspace selection problem for maximizing the projection of a vector of interest. We have introduced several new notions of elemental curvatures, upper bounded by functions of principal angle. We then derived explicit lower bounds for the performance of forward regression and orthogonal matching pursuit in the cases of uniform and non-uniform matroids. Moreover, we showed that if the elements in the ground sets are mutually orthogonal, then these algorithms are essentially optimal under the uniform matroid constraint and they achieve at least $1/2$ approximations of the optimal solution under the non-uniform matroid constraint.

\appendices

\section{Proof of Theorem~\ref{thm2}}
\label{app1}
\begin{IEEEproof}
First consider a subset $E$ of $X$, and two elements $s$ and $t$ in the set $X\setminus E$. we know that $|\langle s| t\rangle| \leq \cos\phi$ by definition of the principal angle. We decompose the two elements into parallel and orthogonal components with respect to $\text{span}(E)$. Let us assume that $\phi_1$ and $\phi_2$ are the angles between $s$, $t$ and $\text{span}(E)$, respectively, then we have
\begin{align*}
&s=\cos \phi_1 \bar s+\sin \phi_1 s^{\perp},\\
&t=\cos \phi_2 \bar t+\sin \phi_2 t^{\perp}.
\end{align*}

We know that
\begin{align*}
\langle s| t\rangle&=\langle \cos \phi_1 \bar s+\sin \phi_1 s^{\perp}| \cos \phi_1 \bar t+\sin \phi_1 t^{\perp}\rangle\\
&=\cos \phi_1 \cos\phi_2 \langle \bar s| \bar t\rangle +\sin\phi_1\sin\phi_2 \langle s^{\perp}| t^{\perp}\rangle.
\end{align*}
Therefore,
\begin{align}\label{ine}
\nonumber
 |\langle s^{\perp}| t^{\perp}\rangle |&=\left| \frac{\langle s| t\rangle-\cos \phi_1 \cos\phi_2 \langle \bar s| \bar t\rangle}{\sin\phi_1\sin\phi_2 } \right|\\
 &\leq \frac{\cos\phi+\cos^2\phi}{\sin^2 \phi}.
\end{align}

For the numerator and denominator in the definitions of curvature, using Pythagoras' theorem, it is easy to show that \[
\|\mathcal P_\eta(E\cup\{t\})\|^2-\|\mathcal P_\eta(E)\|^2 =\| \mathcal P_{\eta} (\{t^{\perp}\})\|^2 = \langle \eta| t^{\perp} \rangle^2, \]
and
 \[
 \|\mathcal P_\eta(E\cup\{s,t\})\|^2-\|\mathcal P_\eta(E\cup\{s\})\|^2 =\| \mathcal P_{\eta} (\{\hat t^{\perp}\})\|^2 = \langle \eta| \hat t^{\perp} \rangle^2, \]
where $\hat t^{\perp}$ denotes the orthonormal component of $t$ with respect to $\text{span}(E\cup \{s\})$. By the Gram--Schmidt process, we know that
\[
\hat t^{\perp}=\frac{t^{\perp}- \langle t^{\perp}| s^{\perp} \rangle s^{\perp} }{\sqrt{1-\langle t^{\perp}| s^{\perp} \rangle^2}}.
\]
 Therefore, we obtain
\[
 \langle \eta| \hat t^{\perp} \rangle = \frac{\langle \eta| t^{\perp} \rangle  - \langle t^{\perp}| s^{\perp} \rangle  \langle \eta| s^{\perp} \rangle }{\sqrt{1-\langle t^{\perp}| s^{\perp} \rangle^2}}.
\]
Hence, using \eqref{ine} we can provide an upper bound of the forward elemental curvature using $\phi$:
\[
\hat \kappa  \leq \frac{(1+\langle t^{\perp}| s^{\perp} \rangle)^2 }{{1-\langle t^{\perp}| s^{\perp} \rangle^2}} ={\frac{1+\langle t^{\perp}| s^{\perp} \rangle }{1-\langle t^{\perp}| s^{\perp} \rangle } }\leq {\frac{1}{1-2\cos \phi}}.
\]
Using a similar argument, we can provide an upper bound for the backward elemental curvature with the same form. The proof is complete.
\end{IEEEproof}

\section{Proof of Theorem~\ref{thm3}}\label{app2}
\begin{IEEEproof}
For any $M,N\in  I$ and $|M|\leq K$ and $|N|=K$, let $J=(M\cup N)\setminus M =\{j_1,\ldots, j_r\}$ where $r\leq |N|$. We can permute the elements in $J$ such that the elements are ordered to use the forward elemental curvature. More specifically, let
\[\bar j_i=\argmin_{j\in J\setminus \{\bar j_1,\ldots, \bar j_{i-1}\}}\| \mathcal P_{\eta}(\{j^{\perp}(M \cup\{\bar j_1,\ldots, \bar j_{i-1}\})\})\|,\] where $j^{\perp}(M \cup\{\bar j_1,\ldots, \bar j_{i-1}\})$ denotes the normalized orthogonal component of $j$ with respect to $\text{span}(M \cup\{\bar j_1,\ldots, \bar j_{i-1}\})$. Using the definition of forward elemental curvature, we have
\begin{align*}
&f(M\cup  N)-f(M)\\
&=\sum_{i=1}^{r} (f(M\cup\{\bar j_1, \ldots, \bar j_i\})
-f(M\cup\{\bar j_1, \ldots, \bar j_{i-1}\})) \\
&\leq\sum_{i=1}^{r} \hat \kappa^{i-1}( f(M\cup\{\bar j_i\})-f(M))
\end{align*}
Therefore, there exists $\bar j\in X$ such that
\begin{align*}
&f(M\cup  N)-f(M)\\
&\leq \sum_{i=1}^{|N|} \hat \kappa^{i-1} ( f(M\cup\{\bar  j\})-f(M))\\
&=\sum_{i=1}^{|N|} \hat \kappa^{i-1}( f(M\cup\{\bar  j\})-f(M)).
\end{align*}
 We use $G_k$ to denote the forward regression solution with cardinality $k$ and $\text{OPT}$ to denote the optimal solution.
Using the properties of the forward regression algorithm and the monotone property, we have
\begin{align*}
&f(G_i)-f(G_{i-1}) \\
&\geq   \frac{1}{\sum_{i=1}^{K} \hat \kappa^{i-1}} (f(G_{i-1}\cup \text{OPT})-f(G_{i-1}))\\
&\geq  \frac{1}{\sum_{i=1}^{K} \hat \kappa^{i-1}}(f(\text{OPT})-f(G_{i-1})).
\end{align*}
Therefore, by recursion, we have
\begin{align*}
f(G_K)&\geq \frac{1}{\sum_{i=1}^{K} \hat \kappa^{i-1}} f(\text{OPT})+(1-\frac{1}{\sum_{i=1}^{K} \hat \kappa^{i-1}})f(G_{K-1}) \\
&= \frac{1}{\sum_{i=1}^{K} \hat \kappa^{i-1}}f(\text{OPT})\sum_{i=0}^{K-1}  (1-\frac{1}{\sum_{i=1}^{K} \hat \kappa^{i-1}})^i \\
&=f(\text{OPT})\left(1-(1-\frac{1}{\sum_{i=1}^{K} \hat \kappa^{i-1}})^{K}\right).
\end{align*}

Using a similar argument, we can show that
\begin{align*}
f(G_K)&\geq f(\text{OPT})\left(1-(1-\frac{1}{\sum_{i=1}^{K} \bar \kappa^{i-1}})^{K}\right).
\end{align*}
Combining these two inequalities, the proof is complete.

\end{IEEEproof}

\section{Proof of Theorem~\ref{thm5}}
\label{app4}
\begin{IEEEproof}
For any $M,N\in  I$ and $|M|\leq K$ and $|N|=K$, let $J=(M\cup N)\setminus M =\{j_1,\ldots, j_r\}$ where $r\leq |N|$. We can permute the elements in $J$ such that the elements are ordered to use the forward elemental curvature. More specifically, let
\[\bar j_i=\argmin_{j\in J\setminus \{\bar j_1,\ldots, \bar j_{i-1}\}}\| \mathcal P_{\eta}(\{j^{\perp}(M \cup\{\bar j_1,\ldots, \bar j_{i-1}\})\})\|,\] where $j^{\perp}(M \cup\{\bar j_1,\ldots, \bar j_{i-1}\})$ denotes the normalized orthogonal component of $j$ with respect the $\text{span}(M \cup\{\bar j_1,\ldots, \bar j_{i-1}\})$. Using the definition of forward elemental curvature, we have
\begin{align*}
&f(M\cup  N)-f(M)\\
&=\sum_{i=1}^{r} (f(M\cup\{\bar j_1, \ldots, \bar j_i\})
-f(M\cup\{\bar j_1, \ldots, \bar j_{i-1}\})) \\
&\leq\sum_{i=1}^{r} \hat \kappa^{i-1}( f(M\cup\{\bar j_i\})-f(M))
\end{align*}
Therefore, there exists $\bar j\in X$ such that
\begin{align*}
&f(M\cup  N)-f(M)\\
&\leq \sum_{i=1}^{|N|} \hat \kappa^{i-1} ( f(M\cup\{\bar  j\})-f(M))\\
&=\sum_{i=1}^{|N|} \hat \kappa^{i-1}( f(M\cup\{\bar  j\})-f(M)).
\end{align*}
Using a similar argument, we can show that
\begin{align*}
&f(M\cup  N)-f(M)\\
&\leq \sum_{i=1}^{|N|} \bar \kappa^{i-1}( f(M\cup\{\bar  j\})-f(M)).
\end{align*}

Using the properties of the forward regression algorithm, the monotone property, and \eqref{eqcomp}, we have
\begin{align*}
&f(T_i) -f(T_{i-1})\\
&\geq \sin^2 \phi(f(T_{i-1}\cup \{g^*\})-f(T_{i-1})) \\
&\geq   \frac{\sin^2 \phi}{\hat {K}} (f(T_{i-1}\cup \text{OPT})-f(T_{i-1}))\\
&\geq  \frac{\sin^2 \phi}{\hat {K}}(f(\text{OPT})-f(T_{i-1})).
\end{align*}
Therefore, by recursion, we have
\begin{align*}
f(G_K)&\geq \frac{\sin^2 \phi}{\hat K} f(\text{OPT})+\left(1-\frac{\sin^2 \phi}{\hat K}\right)f(G_{K-1}) \\
&= \frac{\sin^2 \phi}{\hat K}f(\text{OPT})\sum_{i=0}^{K-1}  \left(1-\frac{\sin^2 \phi}{\hat K}\right)^i \\
&=f(\text{OPT})\left(1-\left(1-\frac{\sin^2 \phi}{\hat K}\right)^{K}\right).
\end{align*}

\end{IEEEproof}

\section{Proof of Thorem~\ref{thm4}} \label{app3}
\begin{IEEEproof}
We use a similar approach as that of the proof of Theorem~\ref{thm3}. Let $G_i=\{g_1,\ldots, g_i\}$ where $g_j$ denotes the element added in the forward regression algorithm at stage $j$.
Let $\text{OPT} =\{o_1,\ldots, o_K\}$ and assume that the elements are already reordered such that we can use the forward elemental curvature.
We know that
\begin{align*}
&f(G_K\cup \text{OPT})  -f(G_K) \\
&\leq \sum_{i=1}^{K} \hat \kappa^{i-1} ( f(G_K\cup\{ o_i\})-f(G_K))\\
&\leq \begin{cases}  \sum_{i=1}^{K} (f(G_{K}\cup \{ o_i\})-f(G_{K})), & \mbox{if } \hat \kappa\leq 1 \\ \hat \kappa^{K-1} \sum_{i=1}^{K} (f(G_{K}\cup \{o_i\})-f(G_{K})), & \mbox{if } \hat \kappa> 1. \end{cases}
\end{align*}

From Lemma~\ref{lemma}, we know that $\text{OPT}$ can be ordered into $\{\hat o_1,\ldots, \hat o_K\}$, such that
\[
f(G_{i-1}\cup \{\hat o_i\})-f(G_{i-1}) \leq f(G_i)-f(G_{i-1}),
\]
for $i=1,\ldots, K$.
Moreover, we know that $G_{i-2}\cup \{g_{i}\} \in I$ because of the hereditary property of the matroid. Moreover, we know that the projection of $\eta$ gains more by adding $g_{i-1}$ than $g_i$ at stage $i-1$ by the property of the forward regression algorithm. Using Lemmas~\ref{lem1} and \ref{lem2} and the definitions of forward and backward elemental curvatures, we obtain \eqref{long1}.
\begin{figure*}[bt]
\normalsize
\begin{align}
 f(G_{K}\cup \{ \hat o_i\})-f(G_{K})   \leq \begin{cases} \hat \kappa (f(G_{K-1}\cup \{\hat o_i\})-f(G_{K-1})), & \mbox{if } \|\mathcal P_{\eta}(\hat o_i^\perp) \| \geq \|\mathcal P_{\eta}(\hat g_K^\perp) \|  \\  \bar \kappa(f(G_{K})-f(G_{K-1})), & \mbox{if } \|\mathcal P_{\eta}(\hat o_i^\perp) \| \leq \|\mathcal P_{\eta}(\hat g_K^\perp) \| . \end{cases}
\label{long1}
\end{align}
\hrulefill
\vspace*{4pt}
\end{figure*}
Therefore, by recursion we have
\begin{align*}
 &f(G_{K}\cup \{ \hat o_i\})-f(G_{K})   \\
&\leq \max(\hat \kappa, \bar \kappa)^{K-i+1} ( f(G_i)-f(G_{i-1})),
\end{align*}
for $i=1,\ldots, K$.
Hence, we obtain
\begin{align*}
 &\quad \sum_{i=1}^{K} (f(G_{K}\cup \{ \hat o_i\})-f(G_{K}))   \\
 &\leq \sum_{i=1}^{K} \max(\hat \kappa, \bar \kappa)^{K-i+1} (f(G_{i})-f(G_{i-1}))  \\
&\leq \begin{cases} \max(\hat \kappa, \bar \kappa) f(G_K), & \mbox{if } \max(\hat \kappa, \bar \kappa)\leq 1 \\ \max(\hat \kappa, \bar \kappa)^{K}f(G_K), & \mbox{if } \max(\hat \kappa, \bar \kappa)> 1. \end{cases}
\end{align*}
Therefore, we have
\begin{align*}
f(\text{OPT})\leq (1+a(\hat \kappa,\bar \kappa)b(\hat \kappa)) f(G_K),
\end{align*}
where $a(\hat \kappa, \bar \kappa)=\max(\hat \kappa, \bar \kappa)$ if $\max(\hat \kappa, \bar \kappa) \leq 1$ and $a(\hat \kappa, \bar \kappa)=\max(\hat \kappa, \bar \kappa)^{K}$ if $\max(\hat \kappa, \bar \kappa) >1$; $b(\hat \kappa)=\hat \kappa^{K-1}$ if $\hat \kappa >1$ and $b(\hat \kappa)=1$ otherwise.
\end{IEEEproof}

\section{Proof of Theorem~\ref{thm6}}
\label{app5}
\begin{IEEEproof}
Let $\text{OPT}=\{o_1,\ldots, o_K\}$ be ordered such that the elemental forward curvature can be used. We know that
\begin{align*}
&f(T_K\cup \text{OPT})  -f(T_K) \\
&\leq \begin{cases}  \sum_{i=1}^{K} (f(T_{K}\cup \{ o_i\})-f(T_{K})), & \mbox{if } \hat \kappa\leq 1 \\ \hat \kappa^{K-1} \sum_{i=1}^{K} (f(T_{K}\cup \{o_i\})-f(T_{K})), & \mbox{if } \hat \kappa> 1. \end{cases}
\end{align*}

Using Lemma 1 and \eqref{eqcomp}, we know that $\text{OPT}$ can be ordered as $\{\hat o_1,\ldots, \hat o_K\}$, such that
\begin{align*}
&f(T_{i-1}\cup \{\hat o_i\})-f(T_{i-1})\\
& \leq f(T_{i-1}\cup\{g^*\})-f(T_{i-1})\\
& \leq \frac{f(T_i) -f(T_{i-1})}{\sin^2 \phi},
\end{align*}
for $i=1,\ldots, K$. Next we state a lemma and its proof that we will use.

\begin{Lemma}
  For $i=1,2,\ldots, K$, we have
\[f(T_i) -f(T_{i-1}) \leq  \tilde\kappa (f(T_{i-1}) -f(T_{i-2})).
\]
\end{Lemma}
\emph{Proof of Lemma 3}: 
For $i=1,\ldots,K$, let $T_i=\{t_1,\ldots, t_i\}$ where $t_j$ denotes the element added in the orthogonal matching pursuit algorithm at stage $j$.
We know that $T_{i-2}\cup \{t_{i}\} \in I$ because of the hereditary property of the matroid. Therefore, we have $|\langle \eta^\perp| t_{i-1}\rangle| \geq |\langle \eta^\perp| t_{i}\rangle|$ by the property of orthogonal matching pursuit. Then, by the definition of the OMP elemental curvature, we obtain the inequality in the lemma.

By the definitions of forward and backward elemental curvatures, we obtain \eqref{long2}.
\begin{figure*}[]
\normalsize
\begin{align}
\label{long2}
 f(T_{K}\cup \{ \hat o_i\})-f(T_{K})   \leq \begin{cases} \hat \kappa (f(T_{K-1}\cup \{\hat o_i\})-f(T_{K-1})), & \mbox{if } \|\mathcal P_{\eta}(\hat o_i^\perp) \| \geq \|\mathcal P_{\eta}(\hat t_K^\perp) \|  \\  \bar \kappa(f(T_{K})-f(T_{K-1})), & \mbox{if } \|\mathcal P_{\eta}(\hat o_i^\perp) \| \leq \|\mathcal P_{\eta}(\hat t_K^\perp) \| . \end{cases}
\end{align}
\hrulefill
\vspace*{4pt}
\end{figure*}
Therefore, by Lemma 3 and recursion, we have
\begin{align*}
 &f(T_{K}\cup \{ \hat o_i\})-f(T_{K})   \\
&\leq \frac{\max(\hat \kappa, \bar \kappa, \tilde \kappa)^{K-i+1}}{\sin^2\phi} ( f(T_i)-f(T_{i-1})),
\end{align*}
for $i=1,\ldots, K$.

Therefore, we have
\begin{align*}
 &\quad \sum_{i=1}^{K} (f(T_{K}\cup \{\hat o_i\})-f(T_{K}))   \\
&=\sum_{i=1}^{K} \frac{ \max(\hat \kappa, \bar \kappa,\tilde \kappa)^{K-i+1} }{\sin^2\phi} (f(T_{i})-f(T_{i-1}))  \\
&\leq \begin{cases} \sin^{-2} \phi \max(\hat \kappa, \bar \kappa, \tilde \kappa) f(T_K), & \mbox{if } \max(\hat \kappa, \bar \kappa,\tilde \kappa)\leq 1 \\  \sin^{-2} \phi \max(\hat \kappa, \bar \kappa, \tilde \kappa)^{K}f(T_K), & \mbox{if } \max(\hat \kappa, \bar \kappa, \tilde \kappa)> 1. \end{cases}
\end{align*}
Therefore, we have
\begin{align*}
f(\text{OPT})\leq (1+ \sin^{-2} \phi a(\hat \kappa,\bar \kappa,\tilde \kappa)b(\hat \kappa)) f(T_K),
\end{align*}
where $a(\hat \kappa, \bar \kappa,\tilde \kappa)=\max(\hat \kappa, \bar \kappa ,\tilde \kappa)$ if $\max(\hat \kappa, \bar \kappa, \tilde \kappa) \leq 1$ and $a(\hat \kappa, \bar \kappa,\tilde \kappa)=\max(\hat \kappa, \bar \kappa,\tilde \kappa)^{K}$ if $\max(\hat \kappa, \bar \kappa,\tilde\kappa) >1$; $b(\hat \kappa)=\hat \kappa^{K-1}$ if $\hat \kappa >1$ and $b(\hat \kappa)=1$ otherwise.

\end{IEEEproof}

\ifCLASSOPTIONcaptionsoff
  \newpage
\fi

\bibliographystyle{IEEEbib}

\begin{thebibliography}{10}

\bibitem{scharf1991statistical}
L.~L. Scharf, ``Statistical signal processing,'' Addison-Wesley, 1991.


\bibitem{lehmann2003theory}
E. L. Lehmann and G. Casella, ``Theory of point estimation,'' Springer Texts in Statistics, 2nd edition, Springer 2003.

\bibitem{mcwhorter1993properties}
L. T. McWhorter and L. L. Scharf, ``Properties of quadratic covariance bounds,'' in \emph{Proceedings of 27th Asilomar Conference on Signals, Systems and Computers,} pp. 1176--1180, vol. 2, Nov. 1993.
\bibitem{todros2010general}
K. Todros, and J. Tabrikian, ``General classes of performance lower bounds for parameter estimation --- part I: non-Bayesian bounds for unbiased estimators,'' \emph{IEEE Transactions on Information Theory}, vol. 56, no. 10, pp. 5045--5063, Oct. 2010.

\bibitem{rao1945information}
C. R. Rao, ``Information and accuracy attainable in the estimation of statistical parameters,''  \emph{Bull. Calcutta Math. Soc.,}  vol. 37,  pp. 81--91, 1945.

\bibitem{bh1947on}
A. Bhattacharyya,  ``On some analogous of the amount of information and their use in statistical estimation,''  \emph{Shankya},  vol. 8,  no. 3,  pp. 201--218 1947.

\bibitem{barankin1946locally}
E. W. Barankin  ``Locally best unbiased estimates,'' \emph{ Ann. Math. Stat.,}  vol. 20,  pp. 477--501, 1946.

\bibitem{hero1997recursive}
A. O. Hero, M. Usman, A. C. Sauve, and J. A. Fessler, ``Recursive algorithms for computing the Cramer-Rao bound,''  \emph{IEEE Transactions on Signal Processing,} vol. 45, no. 3, pp. 803--807, Mar. 1997.

\bibitem{williams2007information}
J. L. Williams,  ``Information theoretic sensor management,'' Ph.D. thesis, MIT, 2007.


\bibitem{krause2007near}
A.~Krause and C.~Guestrin, ``Near-optimal observation selection using
  submodular functions,'' in \emph{Proceedings of National Conference on Artificial
  Intelligence}, vol.~22, no.~2, Vancouver, British Columbia, Canada, Jul. 2007, pp. 1650--1654.

\bibitem{shamaiah2010greedy}
M.~Shamaiah, S.~Banerjee, and H.~Vikalo, ``Greedy sensor selection: Leveraging
  submodularity,'' in \emph{Proceedings of 49th IEEE Conference on Decision and Control}, Atlanta, GA,
Dec. 2010, pp. 2572--2577.



\bibitem{miller2002subset}
A. Miller, ``Subset selection in regression,'' Chapman and Hall, 2nd edition, 2002.


\bibitem{candes2006compressive}
E. J. Candes, ``Compressive sampling,'' in \emph{Proceedings of International Congress Math.,} 2006, vol. 3, pp. 1433--1452.

\bibitem{donoho2006compressed}
D. L. Donoho, ``Compressed sensing,'' \emph{IEEE Transactions on Information Theory,} vol. 52, no. 4, pp. 1289--1306, 2006.

\bibitem{baraniuk2007compresive}
R. G. Baraniuk, ``Compressive sensing,'' \emph{IEEE Signal Processing Magazine,} vol. 24, no. 4, pp. 118 –121, Jul. 2007.

\bibitem{eldar2012compressed}
Y. C. Eldar and G. Kutyniok, ``Compressed sensing: Theory and applications,'' vol. 95. Cambridge, U.K.: Cambridge Univ. Press, 2012.

\bibitem{xu2007efficient}
W. Xu and B. Hassibi, ``Efficient compressive sensing with deterministic guarantees using expander graphs,'' in \emph{Proceedings of IEEE Information Theory Workshop}, pp. 414--419, Sept. 2007.

\bibitem{chi2011sensitivity}
Y. Chi, L. L. Scharf, A. Pezeshki, and A. R. Calderbank, ``Sensitivity to basis mismatch in compressed sensing,'' \emph{IEEE Trans. Signal Processing,} vol. 59, no. 5, pp. 2182--2195, May 2011.


\bibitem{zahedi2013adaptive}
R. Zahedi, L. W. Krakow, E. K. P. Chong, and A. Pezeshki, ``Adaptive estimation of time-varying sparse signals,'' \emph{IEEE Access}, vol. 1, pp. 449--464, 2013.


\bibitem{tropp2004greed}
J. Tropp, ``Greed is good: algorithmic results for sparse approximation,'' \emph{IEEE Transaction on Information Theory,} vol. 50, pp. 2231--2242, 2004.


\bibitem{chi2012coherence}
Y. Chi and R. Calderbank, ``Coherence-based performance guarantees for orthogonal matching pursuit,'' in \emph{Proceedings of
 Allerton Conference on Control, Communications and Computing (Allerton)}, 2012.

\bibitem{kunis2007random}
 S. Kunis and H. Rauhut, ``Random sampling of sparse trigonometric
polynomials, ii. orthogonal matching pursuit versus basis pursuit,'' \emph{Foundations of Computational Mathematics,} vol. 8, no. 6, pp. 737--763, 2007.

\bibitem{cai2011orthognal}
T. Cai and L. Wang, ``Orthogonal matching pursuit for sparse signal recovery with noise,'' \emph{IEEE Transactions on Information Theory}, vol. 57, no. 7, pp. 1--26, 2011.

\bibitem{davenport2010analysis}
M. A. Davenport and M. B. Wakin, ``Analysis of orthogonal matching pursuit using the restricted isometry property,'' \emph{IEEE Transactions on Information Theory}, vol. 56, no. 9, pp. 4395--4401, Sep. 2010.

\bibitem{braunnewinfo}
G. Braun, S. Pokutta, and Y. Xie, ``Info-Greedy sequential adaptive compressed sensing,'' to appear, \emph{IEEE Journal Selected Topics in Signal Processing.}

\bibitem{liu2012the}
E. Liu and V. N. Temlyakov, ``The orthogonal super greedy algorithm and applications in compressed sensing,'' \emph{IEEE Transaction Information Theory}, Vol. 58, no. 4, pp. 2040--2047, 2012.

\bibitem{candes2006stable}
E. Candes, J. Romberg, and T. Tao  ``Stable signal recovery from incomplete and inaccurate measurements,''  \emph{Comm. Pure Appl. Math.},  vol. 59,  no. 8,  pp. 1207--1223, 2006.

\bibitem{wang2012}
Z.~Wang, W.~Moran, X.~Wang, and Q.~Pan, ``Approximation for maximizing monotone
  non-decreasing set functions with a greedy method,'' \emph{Journal of Combinatorial Optimization}, DOI: 10.1007/s10878-014-9707-3, Jan. 2014.

\bibitem{donoho2006stable}
D. L. Donoho, M. Elad, and V. N. Temlyakov, ``Stable recovery of sparse overcomplete representations in the presence of noise,'' \emph{IEEE Transactions on Information Theory,} vol. 52, no. 1, pp. 6--18, Jan. 2006.


\bibitem{tutte1965lectures}
W.~Tutte, ``Lectures on matroids,'' \emph{J. Res. Nat. Bur. Standards Sect. B},
  vol.~69, no. 1-47, p. 468, 1965.

\bibitem{zhang2013neighbor}
L. Zhang, J. Luo, and D. Guo, ``Neighbor discovery for wireless networks via compressed sensing,'' \emph{Performance Evaluation,} vol. 70, no. 7–8, pp. 457--471, Jul. 2013.

\bibitem{nemhauser1978analysis}
G.~L. Nemhauser, L.~A. Wolsey, and M.~L. Fisher, ``An analysis of
  approximations for maximizing submodular set functions---I,''
  \emph{Mathematical Programming}, vol.~14, no.~1, pp. 265--294, 1978.

\bibitem{fisher1978analysis}
M.~L. Fisher, G.~L. Nemhauser, and L.~A. Wolsey, ``An analysis of
  approximations for maximizing submodular set functions---II,'' in
  \emph{Polyhedral Combinatorics}.\hskip 1em plus 0.5em minus 0.4em\relax
  Springer, 1978, pp. 73--87.

\bibitem{conforti1984submodular}
M.~Conforti and G.~Cornuejols, ``Submodular set functions, matroids and the
  greedy algorithm: tight worst-case bounds and some generalizations of the
  Rado-Edmonds theorem,'' \emph{Discrete Applied Mathematics}, vol.~7, no.~3,
  pp. 251--274, 1984.

\bibitem{vondrak2010submodularity}
J.~Vondr{\'a}k, ``Submodularity and curvature: The optimal solution,''
  \emph{RIMS Kokyuroku Bessatsu B}, vol.~23, pp. 253--266, 2010.


\bibitem{zhang2013cdc}
Z. Zhang, Z. Wang, E. K. P. Chong, A. Pezeshki, and W. Moran, ``Near optimality of greedy strategies for string submodular functions with forward and backward curvature constraints," in \emph{Proceedings of the 52nd IEEE Conference on Decision and Control}, Florence, Italy, December 10--13, 2013, pp. 5156--5161.

\bibitem{zhang2013near}
Z.~Zhang, E.~K. P. Chong, A.~Pezeshki, and W.~Moran, ``String submodular functions with curvature constraints,'' to appear in \emph{IEEE Transaction on Automatic Control}, 2015.

\bibitem{liu2014cdc}
Y. Liu, E. K. P. Chong, A. Pezeshki, and W. Moran, ``Bounds for general approximate dynamic programming based on string submodularity and curvature," in \emph{Proceedings of the 53rd IEEE Conference on Decision and Control}, Los Angeles, CA, December 15--17, 2014, pp. 6653--6658.



\bibitem{cevher2011greedy}
V.~Cevher and A.~Krause, ``Greedy dictionary selection for sparse
  representation,'' \emph{IEEE Journal of Selected Topics in Signal Processing,}, vol.~5, no.~5, pp. 979--988, 2011.

\bibitem{das2011}
A.~Das and D.~Kempe, ``Submodular meets spectral: Greedy algorithms for subset
  selection, sparse approximation and dictionary selection,'' in \emph{Proceedings of 28th International Conference on Machine Learning}, Bellevue, WA USA 2011.


\end{thebibliography}

\end{document}